# Is the High-Energy Emission from Centaurus A Compton-Scattered Jet Radiation?

J. G. Skibo[1], C. D. Dermer and R. L. Kinzer

E. O. Hulburt Center for Space Research

Naval Research Laboratory

Code 7650, Washington, DC 20375

## ABSTRACT

We consider whether the hard X-ray and soft gamma-ray emission from Centaurus A is beamed radiation from the active nucleus which is Compton-scattered into our line-of-sight. We derive the spectrum and degree of polarization of scattered radiation when incident beamed radiation is scattered from a cold ($kT << m_e c^2$) electron cloud moving with bulk relativistic motion along the jet axis, and calculate results for an unpolarized, highly-beamed incident power-law photon source. The spectra of the scattered radiation exhibit a cut off at gamma-ray energies due to electron recoil. The cut off energy depends on the observer's viewing angle and the bulk Lorentz factor of the scattering medium. We fit the OSSE data from Centaurus A with this model and find that if the scatterers are not moving relativistically, then the angle the jet makes with respect to our line-of-sight is $61° \pm 5°$. We predict a high degree of polarization of the scattered radiation below $\sim 300$ keV. Future measurements with X-ray and gamma-ray polarimeters could be used to constrain or rule out such a scenario.

*Subject headings:* galaxies: individual — galaxies: active — galaxies: jets — gamma rays: theory — radiation mechanisms: nonthermal — polarization

[1] NRC/NRL Resident Research Associate

# 1. INTRODUCTION

Centaurus A (NGC 5128), at a distance of 3-5 Mpc (z=0.0008), is the brightest extragalactic source at photon energies $\epsilon \sim 100$ keV and one of the nearest active galaxies (see Ebneter & Balick 1983 for a review). The radio emission from Cen A displays a twin-jet structure with radio luminosity $\sim 10^{40}$ ergs s$^{-1}$, suggesting that Cen A is a Fanaroff-Riley Class 1 (FR1) radio galaxy (Fanaroff & Riley 1974). If FR1 galaxies are the parent population of BL Lac objects, as proposed in one version (Browne 1989; Padovani & Urry 1990) of the AGN unification scenario (see Antonucci 1993), Cen A would be classified as a BL Lac object if its jet were coincident with our line-of-sight. Evidence for jet emission in Cen A comes from observations of a one-sided X-ray jet near the nucleus (Feigelson et al. 1981) and, more recently, from observations (Morganti et al. 1991, 1992) of emission line filaments aligned along the direction of the X-ray jet and coincident with the radio jet out to distances $\approx 10$-20 kpc from the nucleus. From measurements of [OIII] line emission, Morganti et al. conclude that the flux of the ionizing radiation in the 2-6 keV band is $\approx 200$ times greater in the jet direction than along our line-of-sight, consistent with the hypothesis that the emission is intrinsically beamed rather than collimated by obscuring material.

Cen A has recently been observed at hard X-ray and gamma-ray energies with the Oriented Scintillation Spectrometer Experiment (OSSE) on the *Compton Gamma Ray Observatory* (CGRO; see Johnson et al. 1993, Kinzer et al. 1994). This instrument operates in the energy range 50 keV $\lesssim \epsilon \lesssim$ 10 MeV. The spectra observed from Cen A are much harder than typical Seyfert spectra (Maisack et al. 1993; Johnson et al. 1994), but Cen A has not been detected at $> 100$ MeV energies (R. C. Hartman, private communication, 1993), a range where many blazars show strong gamma-ray emission (Hartman et al. 1992; Fichtel et al. 1993). In fact, Cen A is only weakly detected with the Compton Telescope on CGRO in the 0.75 - 1.0 MeV energy band by combining data from 3 observing periods (Collmar et al. 1993). A broken power-law gives an acceptable fit to the OSSE Cen A data (Kinzer et al. 1994), but provides no physical basis for the spectral softening in the hard X-ray/soft gamma-ray regime.

In this *Letter*, we examine whether the high-energy emission from Cen A is scattered jet radiation, since Compton scattering of beamed radiation can produce cut offs in photon spectra (Dermer 1993) due to the kinematic recoil of electrons. The precise value of the observed cut off energy is a function of the scattering angle and bulk motion of the electrons. Thus an incident power-law beam of photons will, after scattering, retain its power-law form in the Thompson regime ($\epsilon \ll m_e c^2$ in the rest frame of the electron), but will be cut off above a certain energy in the Klein-Nishina regime ($\epsilon \gg m_e c^2$). In addition, the scattered radiation can be highly polarized, depending on the scattering angle and initial photon polarization.

## 2. ANALYSIS

Consider the Compton scattering of an unpolarized beam of gamma ray photons by a cold ($kT \ll m_e c^2$) cloud moving relativistically along the axis of the beam, which we take to be the positive $z$-axis. Let $\frac{dN_\gamma}{dt d\Omega d\epsilon}(\epsilon, \Omega)$ represent the rate that photons are emitted from the central source (photons s$^{-1}$ sr$^{-1}$ MeV$^{-1}$). In the single scattering approximation ($\tau \ll 1$) the photon flux (photons s$^{-1}$ cm$^{-2}$ MeV$^{-1}$) scattered into the direction $\Omega_0$ with respect to the $z$-axis and received by an observer located at distance $r$ from the scattering cloud is given by the expression

$$\Phi_s(\epsilon_s, \Omega_0) = \frac{1}{r^2 [\gamma(1-\beta\cos\theta_0)]^2} \oint d\Omega^* \frac{N_e^*(\Omega^*)}{|\frac{d\epsilon_s^*}{d\epsilon^*}|} \frac{d\sigma_C}{d\Omega_0^*}(\epsilon^*, \theta_s^*) \frac{dN_\gamma^*}{dt^* d\Omega^* d\epsilon^*}(\epsilon^*, \Omega^*). \qquad (1)$$

Here $N_e^*(\Omega^*)$ is the electron column density in the direction $\Omega^*(\Omega)$ and $\frac{d\sigma_C}{d\Omega_0^*}(\epsilon^*, \theta_s^*)$ is the differential Compton cross section (cm$^2$ ster$^{-1}$) for the scattering of a photon with initial direction $\Omega^*$ and energy $\epsilon^*$ through an angle $\theta_s^*$ into the direction $\Omega_0^*$ with energy $\epsilon_s^*$. The starred symbols represent quantities in the rest frame of the scattering cloud which is moving with velocity $\beta c$ ($\gamma = 1/\sqrt{1-\beta^2}$) along the $z$-axis in the stationary frame.

For a photon of energy $\epsilon$ and direction $(\theta, \phi)$, the relevant transformations are $\phi^* = \phi$, $\cos\theta^* = (\cos\theta - \beta)/(1 - \beta\cos\theta)$, and $\epsilon^* = \gamma(1 - \beta\cos\theta)\epsilon$. In equation (1) the factor $\gamma^{-2}(1 - \beta\cos\theta_0)^{-2}$ relates the scattered emission in the rest frame of the cloud to the emission received by the observer (cf. eq. [4.97b] of Rybicki & Lightman 1979).

The scattered photon energy is given by the well-known formula

$$\epsilon_s^* = \frac{\epsilon^*}{1 + \frac{\epsilon^*}{m_e c^2}(1 - \cos\theta_s^*)}, \tag{2}$$

where $\cos\theta_s^* = \cos\theta^*\cos\theta_0^* + \cos\phi^*\sin\theta^*\sin\theta_0^*$. It can be seen upon inspection of equation (2) that the recoil of the electrons imposes a kinematical cut off in the emergent photon spectrum at an energy in the rest frame of the scatterer given by

$$\epsilon_c^* = \frac{m_e c^2}{1 - \cos\theta_s^*}. \tag{3}$$

In this frame the cut off is only a function of the scattering angle $\theta_s^*$. However, in the stationary frame, this cut off occurs at an energy given by

$$\epsilon_c = \frac{m_e c^2}{\gamma(1 + \beta)(1 - \cos\theta_s)}. \tag{4}$$

Hence, the cut off in the spectrum is both a function of the scattering angle $\theta_s$ and the bulk Lorentz factor $\gamma$ of the scattering cloud.

We consider two cases for the incident photon number intensity: (1) a mono-directional power-law beam along the positive $z$-axis, given by

$$\frac{dN_{\gamma 1}}{dt d\Omega d\epsilon}(\epsilon, \theta) = \frac{1}{2\pi} k_1 \epsilon^{-\alpha} \delta(\cos\theta - 1); \tag{5}$$

and (2) a power-law photon source which is isotropic in a frame moving relativistically along the positive $z$-axis with velocity $\beta_0 c$. The emission in the stationary frame in case 2 is beamed by virtue of relativistic motion and we have

$$\frac{dN_{\gamma 2}}{dt d\Omega d\epsilon}(\epsilon, \theta) = \frac{k_2 \epsilon^{-\alpha}}{[\gamma_0(1 - \beta_0\cos\theta)]^{2+\alpha}}. \tag{6}$$

Synchrotron-self Compton models (e.g., Bloom & Marscher 1993) and models involving the Compton scattering of external photons (Dermer & Schlickeiser 1993; Sikora, Begelman, & Rees 1994) both produce beaming patterns characterized by equation (8). The constants $k_1$ and $k_2$ in equations (5) and (6) are easily normalized to the total jet luminosity or photon number.

In the rest frame of the scattering cloud, the expressions for the received radiation are

$$\frac{dN^*_{\gamma 1}}{dt^* d\Omega^* d\epsilon^*}(\epsilon^*, \theta^*) = \frac{k_1 \epsilon^{*-\alpha} \delta(\cos\theta^* - 1)}{2\pi \left[\gamma(1 + \beta\cos\theta^*)\right]^\alpha}, \tag{7}$$

$$\frac{dN^*_{\gamma 2}}{dt^* d\Omega^* d\epsilon^*}(\epsilon^*, \theta^*) = \{\gamma\gamma_0[1 + (\beta - \beta_0)\cos\theta^* - \beta\beta_0]\}^{-(2+\alpha)} k_2 \epsilon^{*-\alpha}. \tag{8}$$

The scattered photon flux is obtained by substituting equation (7) or (8) into equation (1) and using the expression

$$\frac{d\sigma_C}{d\Omega^*_0}(\epsilon^*, \theta^*_s) = \frac{r_0^2}{2} \left(\frac{\epsilon^*_s}{\epsilon^*}\right)^2 \left(\frac{\epsilon^*_s}{\epsilon^*} + \frac{\epsilon^*}{\epsilon^*_s} - \sin^2\theta^*_s\right) \tag{9}$$

where $r_0 = e^2/m_e c^2 = 2.82 \times 10^{-13}$ cm$^2$ is the classical electron radius (e.g. Rybicki & Lightman 1979). For the mono-directional power law beam (case 1) we have

$$\Phi_{s1}(\epsilon_s, \theta_0) = \frac{N_e r_0^2 k_1 \gamma^\alpha (1-\beta)^\alpha \epsilon^{*-\alpha}}{2r^2 \gamma^2 (1 - \beta\cos\theta_0)^2} \left(\frac{\epsilon^*_s}{\epsilon^*} + \frac{\epsilon^*}{\epsilon^*_s} - \sin^2\theta^*_0\right), \tag{10}$$

where $\epsilon^*$ is obtained by inverting equation (2) with $\theta^*_s = \theta^*_0$. For case 2 the scattered flux is

$$\Phi_{s2}(\epsilon_s, \theta_0) = \frac{N_e r_0^2 k_2}{2r^2 \gamma^2 (1 - \beta\cos\theta_0)^2} \oint d\Omega^* \, \epsilon^{*-\alpha} \left(\frac{\epsilon^*_s}{\epsilon^*} + \frac{\epsilon^*}{\epsilon^*_s} - \sin^2\theta^*_s\right) \\ \times \{\gamma\gamma_0[1 + (\beta - \beta_0)\cos\theta^* - \beta\beta_0]\}^{-(2+\alpha)}, \tag{11}$$

In deriving equation (11), we have assumed that the scattering region has constant electron column density $N_e$ and is much larger than the angular extent of the beam. This result is correct in a regime where the light-travel time through the cloud is small in comparison with the time scale of variation of the central source. A relativistically correct formalism of the time-dependent system is in preparation by the authors.

In Figure 1, we plot the scattered photon spectra multiplied by $\epsilon_s^2$ for various scattering angles and bulk Lorentz factors $\gamma$. The spectral index $\alpha$ of the incident photon spectrum is set equal to 1.6, close to values measured for Cen A at hard X-ray energies (e.g., Baity et al. 1981). The dotted curves represent the flux obtained using equation (10) (case 1) and the solid curves were obtained using equation (11) (case 2). In case 2, we plot the quantity $2m_e c^2 r^2 \Phi_{s2}/N_e r_0^2 k_2$ and take $\gamma_0 = 10$. The normalization of the flux in case 1 was adjusted to make the low-energy fluxes for case 1 and case 2 coincide.

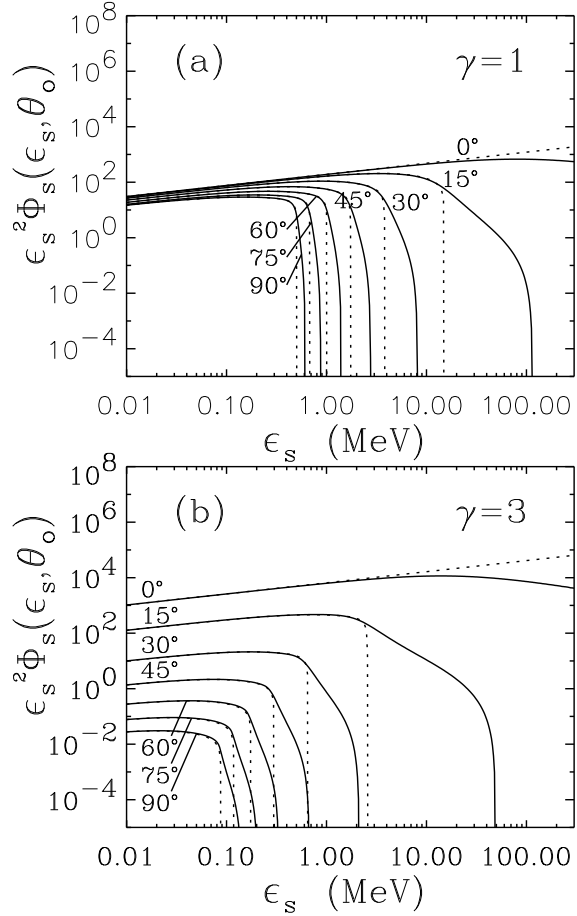

Figure 1. Scattered photon flux multiplied by $\epsilon_s^2$ as a function of energy for various scattering angles. The different panels correspond to bulk Lorentz factors of the scattering cloud of $\gamma = 1, 3$. The dotted (solid) curves represent case 1 (case 2) as described in the text.

For case 1, the mono-directional beam, the high-energy cut offs $\epsilon_c$ are very sharp due to the kinematic constraint expressed by equation (6). At $\epsilon \ll \epsilon_c$, the spectra are power laws with spectral indices equal to $\alpha$. Slight spectral softenings are apparent at $\epsilon \lesssim \epsilon_c$. For case 2, the cut offs are less sharp than in case 1 due to the angular extent of the incident beam. This effect is enhanced for larger bulk Lorentz factors of the scattering cloud. For $\gamma = \gamma_0$, the incident beam becomes isotropic (see eq. [8]).

In general, the scattered radiation will be partially polarized with the direction of the electric field vector perpendicular to the jet axis. The degree of polarization as a function of scattering angle is given by the expression (e.g. McMaster 1961):

$$\Pi(\epsilon_s, \theta_s) = \frac{\sqrt{Q^2 + U^2 + V^2}}{I} = \frac{\sin^2 \theta_s}{\frac{\epsilon_s}{\epsilon} + \frac{\epsilon}{\epsilon_s} - \sin^2 \theta_s}, \qquad (12)$$

where $I$, $Q$, $U$ and $V$ are the Stoke's parameters (e.g. Rybicki & Lightman 1979) of the scattered radiation. Equation (12) is valid for a mono-directional beam. To obtain the degree of polarization of a Compton scattered beam in case 2 it is necessary to average this expression in the starred system over the angular extent of the incident beam weighted by the differential scattering rate $\frac{dN^*_{\gamma_s}}{dt^* d\Omega^*_s d\epsilon^*_s}$, which is proportional to the integrand in equation (11). This follows from the additivity of the Stoke's parameters and the fact that $U = V = 0$ for Compton scattering of an unpolarized beam by unpolarized electrons. In addition, the degree of polarization is a Lorentz invariant quantity in the sense that $\Pi^*(\epsilon^*_s, \theta^*_s) = \Pi(\epsilon_s, \theta_s)$. Hence, we obtain the expression

$$\langle \Pi(\epsilon_s, \theta_0) \rangle = \langle \Pi^*(\epsilon^*_s, \theta^*_0) \rangle = \frac{\oint d\Omega^* \frac{dN^*_{\gamma_s}}{dt^* d\Omega^*_s d\epsilon^*_s} \Pi^*(\epsilon^*_s, \theta^*_s)}{\oint d\Omega^* \frac{dN^*_{\gamma_s}}{dt^* d\Omega^*_s d\epsilon^*_s}}. \qquad (13)$$

In Figure 2, the degree of polarization is plotted as a function of energy for various scattering angles and bulk Lorentz factors. Here, as in Figure 1, we set $\alpha = 1.6$ and $\gamma_0 = 10$. For a stationary scattering cloud (panel a), the degree of polarization is greatest for Thomson scattering ($\epsilon \ll m_e c^2$) through $90°$. The maximum polarization for a cloud with bulk relativistic motion is found in the direction $\cos\theta_s = \beta$, which follows from the invariance of the degree of polarization.

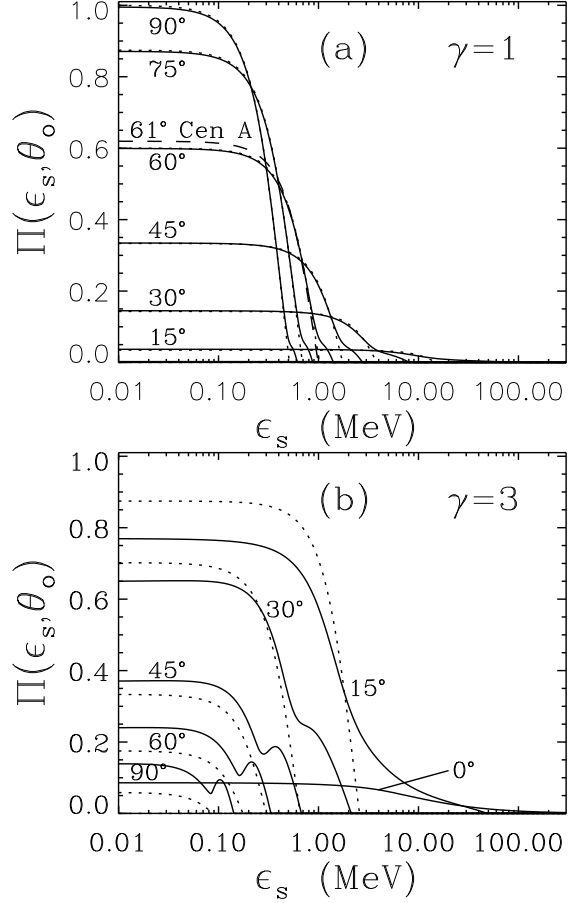

Figure 2. The degree of polarization of the scattered radiation as a function of energy for the various scattering angles. The different panels correspond to bulk Lorentz factors of the scattering cloud of $\gamma = 1, 3$. The dotted (solid) curves represent case 1 (case 2) as described in the text. The dashed curve in panel a is the predicted degree of polarization for Cen A.

The finite angular extent of the beam in case 2 has the effect of increasing the degree of polarization relative to the mono-directional case in the Klein-Nishina regime. This is more pronounced for larger bulk Lorentz factors of the scattering cloud. The bump just below the high-energy cut off in panel (b) is the result of the increasingly limited domain of the angular integration with increasing photon energy imposed by the Klein-Nishina cross section in equation (13).

## 3. RESULTS AND DISCUSSION

We consider the case where the scattering cloud is at rest ($\gamma \cong 1$, $\beta \ll 1$), and fit the spectrum given by equation (10) for a mono-directional beam to the data obtained with OSSE for Cen A during 17-31 October 1991 (Kinzer et al. 1994). This is done by folding the model photon flux through the full OSSE response matrix and conducting the $\chi^2$ search in count space. There are three free parameters in this fit: the overall normalization, $N_e r_0^2 k_1 / 2r^2$; the spectral index of the incident beam, $\alpha$; and the scattering angle, $\theta_0$. The results are shown in Fig. 3 where the curve corresponds to the best fit with a reduced $\chi^2$ of 0.997 for 158 degrees of freedom, giving a probability of 0.496 to obtain this much or more scatter in the data. The parameters for this fit are $\alpha = 1.68 \pm 0.03$, $\theta_0 = 61° \pm 5°$ and $N_e r_0^2 k_1 / 2r^2 = (9.8 \pm 1.3) \times 10^{-4}$ (photons s$^{-1}$ cm$^{-2}$ MeV$^{\alpha-1}$). The polarization of the gamma ray emission is shown as a function of energy for this inclination angle by the dashed curve in panel (a) of Figure 2. The emission is approximately 60% polarized for $\epsilon \lesssim 300$ keV, but the polarization falls rapidly to zero above this energy.

Thus we find that if the scattering cloud is at rest, the jet in Cen A is directed at an angle of $61° \pm 5°$ with respect to our viewing direction. Studies of the distribution of HII regions in NGC 5128 imply that the direction of our line-of-sight is oriented by $73° \pm 3°$ (Graham 1979) or $72° \pm 2°$ (Dufour et al. 1979) with respect to the normal to the plane of the disk of the HII regions and, presumably, Cen A's galaxy. If one can assume that the jet axis is perpendicular to this plane, then the results of these studies are within $2\sigma$ of our inferred angle. If the cloud is moving relativistically, as might be suggested by time variability in the OSSE Cen A data on time scales $\lesssim 0.5$ days (Kinzer et al. 1994), then the deduced angle could be much different. For example, if $\gamma = 2$, then a similar analysis produces a fit with an inclination angle $\approx 35°$. Nonrelativistic scatterers could be provided, however, by material confining the jet, high-energy particles decelerated to nonrelativistic energies by Compton drag or, in the two-flow model (e.g., Sol, Pelletier, & Asséo 1989), by a background electron-proton plasma which supports a pair jet.

We now show that if the X-ray and gamma-ray flux from Cen A is scattered jet radiation, then the measured flux is consistent with optical determinations of the source energetics and the assumption that the scattering cloud is optically thin. Morganti et al. (1991), from their analysis of the [OIII] line emission, conclude that the intensity of ionizing radiation in the jet is at least $10^{54\pm1}$ photons s$^{-1}$ sr$^{-1}$, and is $\sim 200$ times more intense in the jet direction than along our line-of-sight. Using the spectrum derived from their model and equation (5), we obtain $k_1 \cong 3 \times 10^{49}\Omega_b$ photons s$^{-1}$ MeV$^{0.68}$, where $\Omega_b$ is the angular extent of the ionizing beam and $r = 3.7$ Mpc, the value used by Morganti et al. (1991). From the normalization in the fit we obtain an electron column density of $N_e \cong 10^{23}\Omega_b^{-1}$ cm$^{-2}$. This corresponds to a Thompson optical depth $\tau_T \cong 0.07\Omega_b^{-1}$, in agreement with a simple estimate obtained by letting the ratio of scattered flux to the flux of the primary jet radiation $\approx \tau_T \Omega_b/4\pi \approx 1/200$. Hence, the optically-thin approximation is adequate if the opening angle of the beam is $\gtrsim 10°$, and we see that the observed $\gtrsim 1$ keV X-ray flux from Cen A could be entirely scattered jet radiation.

Throughout this analysis we assumed that the polarization of the incident beam was negligible along the jet axis, as suggested by some models where the jet emission is upscatterd disk radiation (e.g., Begelman & Sikora 1987). The polarization properties of blazar jet emission (e.g., Mrk 421) can be directly measured to improve this assumption, but we note that except for specialized scattering geometries, the degree of polarization of the scattered radiation is, in general, greater than that determined assuming an unpolarized incident beam. Hence the degree of polarization we obtain from an initially unpolarized beam probably gives a lower limit. The presence of accretion disk emission from a Seyfert nucleus in Cen A could, however, lower the degree of polarization, but the average spectrum of Seyfert galaxies is very soft at $\epsilon \gtrsim 100$ keV (Johnson et al. 1994).

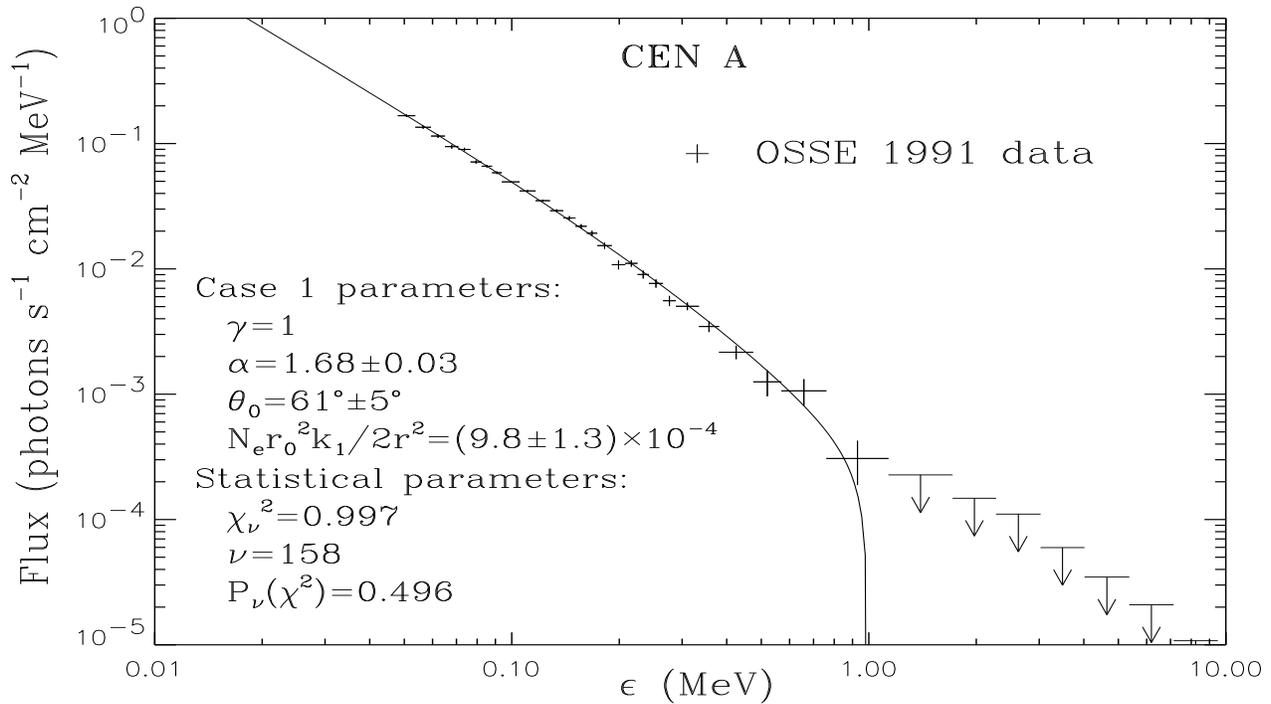

Figure 3. The gamma ray emission from Cen A as detected by OSSE during 17-31 October 1991. The curve represents the best fit Compton scattered spectrum for an ideal beam (Case 1) scattered by a cold stationary ($\gamma = 1$) electron cloud.

## 4. CONCLUSIONS

We have examined whether the hard X-ray and soft gamma-ray emission observed from Cen A could be jet radiation scattered into our line-of-sight by an electron scattering cloud. Models for jet formation in AGN (e.g., Begelman, Blandford, & Rees 1984; Dermer & Schlickeiser 1992) invoke the existence of outflowing plasma, so this process should operate at some level. Making the simplest assumption that the scattering cloud is at rest, we obtain a good fit to the OSSE data for Cen A during the 1991 observations. Moreover, we see that gamma-ray observations provide a new method for the determination of the angle between the jet and viewing directions which does not involve assumptions regarding the orientation of the jet and the plane of the galaxy. This could be crucially important to the interpretation of radio observations of superluminal sources, where the speeds of the jets are inferred indirectly since this angle is not known. Unfortunately, this method is not without its problems: it can only be applied to the brightest sources with the present generation of gamma-ray telescopes. Moreover, the scattering gas may not be cold and could have some bulk relativistic motion which, as we have seen, considerably changes the inferred jet angle. There is, furthermore, no guarantee that the observed radiation is not direct jet radiation nor central source emission. Polarization measurements in the hard X-ray/soft gamma-ray regime would be extremely important in assessing whether the emission from Cen A is direct or reprocessed radiation, since the polarization properties of direct blazar jet emission and scattered radiation should be very different. We encourage hard X-ray and gamma-ray polarization measurements to test these ideas.

*Acknowledgements:* We thank Jim Kurfess for comments on the manuscript, and Kellie McNaron-Brown for assistance with the IGORE software.

## REFERENCES

Antonucci, R. 1993, ARA&A, 31, 473

Baity, W. A. et al. 1981, ApJ, 244, 429

Begelman, M. C., Blandford, R. D. & Rees, M. J. 1984, Rev. Mod. Phys., 56, 1984

Begelman, M. C. & Sikora, M. 1987, ApJ, 322, 650

Bloom, S. D. & Marscher, A. P. 1993, in The Compton Gamma-Ray Observatory, ed. M. Friedlander, N. Gehrels & D. Macomb (New York: AIP), 578

Browne, I. W. A. 1989, in BL Lac Objects, ed. L. Maraschi, T. Maccacaro & M.-H. Ulrich (Springer-Verlag), 401

Collmar, W. et al. 1993, in The Compton Gamma-Ray Observatory, ed. M. Friedlander, N. Gehrels & D. Macomb (New York: AIP), 483

Dermer, C. D. 1993, in Proceedings of the Fifth International Workshop on Neutrino Telescopes, ed. M. Baldo Ceolin, p. 427

Dermer, C. D. & Schlickeiser, R. 1992, Science, 257, 1642

Dermer, C. D. & Schlickeiser, R. 1993, ApJ, 416, 458

Dufour, R. J., van den Berg, S., Harvel, C. A., Martins, D. H., Schiffer III, F. H., Talbot Jr., R. J., Talent, D. L., & Wells, D. C. 1979, AJ, 84, 284

Ebneter, K. & Balick, B. 1983, Publ. Astron, Soc. Pac. 95, 675

Fanaroff, B. L. & Riley, J. M. 1974, MNRAS, 167, 31P

Feigelson, E. D., Schreier, E. J., Delvaille, J. P., Giacconi, R., Grindlay, J.E., & Lightman, A. P. 1981, ApJ, 251, 31

Fichtel, C. E. et al. 1993, in The Compton Gamma-Ray Observatory, ed. M. Friedlander, N. Gehrels & D. Macomb (New York: AIP), 461

Graham, J. A. 1979, ApJ, 232, 60

Hartman, R. C. et al. 1992, ApJ, 385, L1

Johnson W. N. et al. 1993, A&AS, 97, 21

Johnson W. N. et al. 1994, in The Second Compton Symposium, ed. C. E. Fichtel, N. Gehrels & J. Norris (New York: AIP), in press

Kinzer, R. L. et al. 1994, in The Second Compton Symposium, ed. C. E. Fichtel, N. Gehrels & J. Norris (New York: AIP), in press

Maisack, M. et al. 1993, ApJ, 407, L61

McMaster, W. H. 1961, Rev. Mod. Phys., 33, 8

Morganti, R., Robinson, A., Fosbury, R. A. E., diSerego Alighieri, S., Tadhunter, C. N., & Malin, D. F. 1991, MNRAS, 249, 91

Morganti, R., Fosbury, R. A. E., Hook, R. N., Robinson, A., & Tsvetanov, Z. 1992, MNRAS, 256, 1p

Padovani, P. & Urry, C. M. 1990, ApJ, 356, 75

Rybicki, G. B. & Lightman, A. P. 1979, Radiation Processes in Astrophysics, (New York: Wiley)

Sikora, M. Begelman, M. C., & Rees, M. J., 1994, ApJ, in press

Sol, H., Pelletier, G., & Asséo, E. 1989, MNRAS, 237, 411